# Correlating surface energy with adsorption energy by means of intrinsic characteristics of substrates

Bo Li, Xin Li, Wang Gao,* and Qing Jiang*

Key Laboratory of Automobile Materials, Ministry of Education, Department of Materials Science and Engineering, Jilin University 130022, Changchun, China.

**ABSTRACT:** Surface energy is fundamental in controlling surface properties and surface-driven processes like heterogeneous catalysis, as adsorption energy is. It is thus crucial to establish an effective scheme to determine surface energy and its relation with adsorption energy. Herein, we propose a model to quantify the effects of materials' intrinsic characteristics on the material-dependent property and anisotropy of surface energy, based on the period number and group number of bulk atoms, and the valence-electron number, electronegativity and coordination of surface atoms. Our scheme holds for elemental crystals in both solid and liquid phases, body-centered-tetragonal intermetallics, fluorite-structure intermetallics, face-centered-cubic intermetallics, Mg-based surface alloys and semiconductor compounds, which further identifies a quantitative relation between surface energy and adsorption energy and rationalizes the material-dependent error of first-principle methods in calculating the two quantities. This model is predictive with easily accessible parameters and thus allows the rapid screening of materials for targeted properties.

## Introduction

Surface energy is fundamental and dominant in controlling surface structure, reconstruction, roughening, nanoparticles' size, and crystal's shape for solids[1-4]. As surfaces are the region where materials interact with media, surface energy is also of great importance in determining surface-driven processes such as heterogeneous catalysis, gas sensing and biomedical applications[5-8], as adsorption energy is. Thus, it is crucial to establish an effective scheme to determine surface energy and its relation with adsorption energy, particularly by means of the easily accessible intrinsic characteristics of materials. However, the electronic and geometric factors that dictate the change of surface energy from one material to the next (material-dependent property) and control the anisotropy of surface energy, still remain elusive so far. On the other hand, although it is generally assumed a positive correlation between surface energy and adsorption energy[9,10], the quantitative relationship as well as the underlying physical picture still remains debated.

Many (semi-)empirical models have been proposed for determining surface energy of solids. Broken-bond models[11-13] correlate surface energy with the energy of broken chemical bonds but are only applicable into the low-index surfaces of elemental crystals. The Miedma model[14] and Stefan model[15] predict surface energy with the experimental energy of vaporization and are inconvenient for the application in many cases like surface alloys. The Friedel model[16] describes surface energy with the $d$-band width and thus is only applicable into transition metals (TMs) and TM alloys. Although these models have been applied with some success, they are not related to the easily accessible intrinsic properties of materials, perform with limited universality and effectiveness, and can hardly reveal the connection between surface energy and adsorption energy.

First-principle methods have been a workhorse in characterizing surface properties. Compared to experiments, the widely used semi-local functionals underestimate surface energy and overestimate adsorption energy on Pt and Rh, but underestimate both two quantities on Ag and Au, exhibiting an unexpected material-dependent property[9,10,17-20]. This drawback poses an obstacle to the reliable predictions of surface reconstructions. Random phase approximation (RPA)[21] calculations can resolve the dilemma on Pt and Rh and produce improved results, which was attributed to the better description of electronic structure by RPA than by the semi-local functionals[10]. However, RPA calculations are too expensive to be widely used for material screening and still generate material-dependent error in describing surface energy and adsorption energy[10,19]. These methods clearly indicate a strong correlation between surface energy and adsorption energy, however, their numerical characteristics prohibit the understanding of the underlying physical framework. Hence, it is urgently needed to set out from the intrinsic property of materials to understand the correlation between surface energy and adsorption energy.

Here we propose a universal picture to determine surface energy and its correlation with adsorption energy, by using the period number and group number of bulk atoms, and the valence-electron number, electronegativity and coordination number of surface atoms. This model is predictive for a variety of materials covering elemental crystals in both solid and liquid phases, alloys, and semiconductor compounds. Furthermore, our scheme builds a quantitative relation between surface energy and adsorption energy, which allows the estimation of adsorption energy with surface energy and rationalizes the material-dependent error of first-principle methods in calculating surface energy and adsorption energy.

## Results

We study surface energies for 45 elemental crystals, 31 alloys, and 12 compounds including TMs, main-group crystals, Mg-based surface alloys and III-V semiconductors, and cleavage energies for 11 body-centered-tetragonal (AB) intermetallics, 11 fluorite-structure ($A_2B$) intermetallics and 13 face-centered-cubic ($A_3B$) intermetallics. Note that surface energy and cleavage energy denote the energy required to cleave the bulk materials for symmetric and asymmetric terminations respectively. These solids exhibit

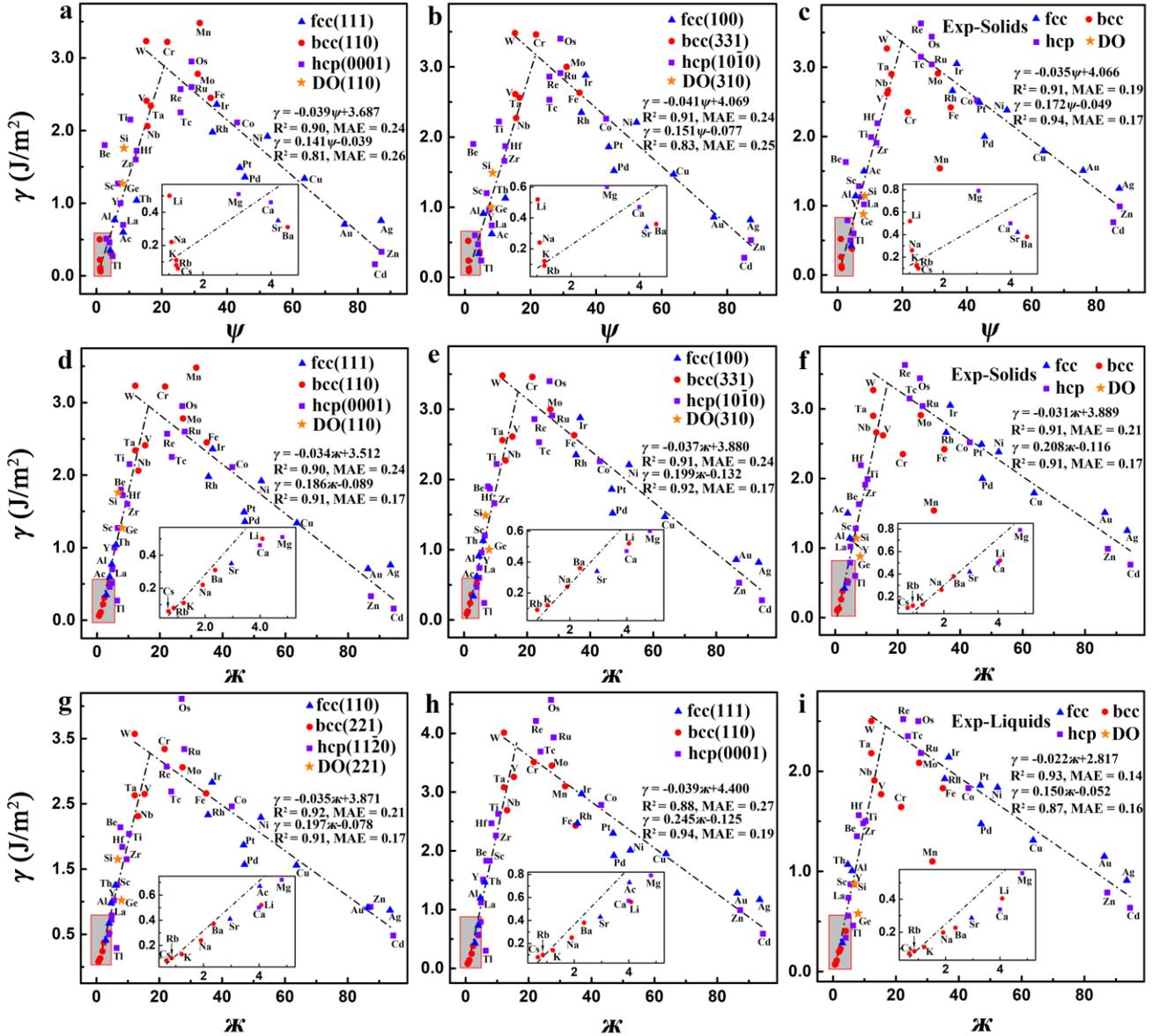

**Figure 1 | Surface energies of elemental crystals against the electronic descriptors ψ and 𝒦 in both solid and liquid phases**. a, b, c, d, e, f, Comparison between ψ and 𝒦 in describing DFT-calculated [with projector augmented wave (PAW) basis set and PBE functional[28]] and experimental surface energies of elemental crystals[14, 28]. g, DFT-calculated surface energies [with PAW basis set and PBE functional[28]] versus 𝒦. h, DFT-calculated surface energies [with the full-charge density (FCD) Green function LMTO technique in the atomic-sphere approximation (ASA)[26]] versus 𝒦. i, Experimental surface energies of elemental crystals in liquid phases versus 𝒦[34–41]. Different orientations of crystals are put together based on the similar $CN/\overline{CN}$ term and the stability of materials.

6 different crystal structures, such as face centered cubic (fcc), close-packed hexagonal (hcp), body centered cubic (bcc), diamond (DO), body centered tetragonal and fluorite structures. The corresponding surfaces contain up to 13 different crystal orientations for elemental crystals, the close-packed surface (CPS) for each liquid phase and III-V semiconductor compounds, 3 different orientations for Mg-based alloys, 12 different orientations for AB intermetallics, 4 different orientations for $A_2B$ intermetallics and 3 different orientations for $A_3B$ intermetallics. Note that the CPSs correspond to fcc(111), bcc(110), hcp(0001), and DO (110) surfaces.

**The material-dependent property of surface energy.** We first attempt to describe surface energy and cleavage energy with surface properties, since it is generally accepted that they are directly correlated with the binding energy of surface atoms[11–13]. The $d$-band model and Muffin-Tin-Orbital theory show that the spatial extent of the metal $d$-orbitals on surfaces is associated with the number of outer electrons[22], indicating that the binding energy of surface atoms likely depends on the number of valence electrons. In addition, Pauling electronegativity ($\chi$) is directly related to the interatomic binding energy. We thus try to describe surface energy and cleavage energy using valence number and electronegativity of surface atoms (which are known to be descriptive for adsorbate-surface binding on TMs[23]), with $\psi = \frac{S_v^2}{\chi^\beta}$, where $S_v$ signifies the valence number with the maximum value 12 (including both the $sp$- and $d$- electrons for TMs). $\beta$ is a parameter determined by the contribution of $d$-orbitals and/or $sp$- orbitals to valence description and electronegativity. The $sp$-orbitals contribution is 100% with $\beta$=1 for main-group crystals (without valence $d$-electrons), while $d$-orbitals and $s$-orbitals have the equal contribution for TMs (each has $\beta = 1/2$). $\beta$ is 1/2 for Ag and Au and is 1 for the other crystals, because Au and Ag exhibit the full-filled $d$-band and the low position of $d$-band center relative to the Fermi levels (-3.56 for Au and -4.40 eV for Ag)[24,25] and bind with other atoms mainly via $sp$ states (although they have valence $d$-electrons). All of the $\psi$ values correspond to elemental crystals are listed in Supplementary Tables 1 and 2. Fig. 1a-c and Supplementary Fig. 1a-c plot the DFT-calculated and experimental surface energy $\gamma$ versus $\psi$ for elemental crystals[14,26–28]. Clearly, $\gamma$ is approximately linearly scaled with $\psi$ in a broken-line behavior. However, $\psi$ fails to elucidate the trends of surface energy for alkaline metals and alkaline-earth metals (see the insets in Fig. 1a-c and Supplementary Fig. 1a-c).

To address this issue, we introduce another two new bulk parameters to describe surface energy and cleavage energy, the period number ($N_p$) and group number ($N_g$) of bulk atoms, since bulk properties are also important for surface stability. The underlying mechanism will be explained in the section of "Understanding and progress of the model". We now propose a new descriptor for describing surface energy and cleavage energy as follows,

$$\mathcal{K} = \left(\frac{N_p}{\overline{N_p}}\right)^{\left(\sqrt{N_g} - \sqrt{\overline{N_g}}\right)} \times \psi = \left(\frac{N_p}{4}\right)^{\left(\sqrt{N_g} - 3\right)} \times \frac{S_o^2}{\chi^\beta} \quad (1)$$

$\overline{N_p}$ and $\overline{N_g}$ are the average period number and average group number for all elements, which are constant 4 and 9. $\mathcal{K}$ is easy to acquire since the involved parameters are available from the periodic table of elements. All of the $\mathcal{K}$ values for elemental crystals are listed in Supplementary Tables 1 and 2.

We now plot the DFT-calculated and experimental surface energy $\gamma$ versus $\mathcal{K}$ for elemental crystals in Fig. 1d-i and Supplementary Figs. 1-3[14,26–28]. Remarkably, $\mathcal{K}$ describes the trends of the surface energy very well for alkaline metals and alkaline-earth metals as well as other elemental crystals. Meanwhile, the description accuracy with $\mathcal{K}$ is overall improved compared to that with $\psi$. These results imply that the period number and group number of bulk atoms, and the valence-electron number and electronegativity of surface atoms together determine surface energy.

The fitted scaling relations of $\gamma$ versus $\mathcal{K}$ are as follows,

$$\gamma = k\mathcal{K} + b, \quad \begin{cases} \mathcal{K} < 17 \\ \mathcal{K} > 17 \end{cases} \quad (2)$$

where $k$ and $b$ are the slope and offset of the scaling relation. We identify that the slope $k$ of the linear relation is approximately -0.035 for crystals with $\mathcal{K} > 17$ and 0.20 for crystals with $\mathcal{K} < 17$. The turning point of the scaling of $\gamma$ versus $\mathcal{K}$ is around Cr and Re that exhibit half-occupied $d$-bands, which can be well understood with the $d$-band occupation. The higher half of $d$-bands corresponds to anti-bonding states and the lower half to bonding states, indicating the most stable bonding at half-occupied $d$-bands[11,26]. Notably, the offset $b$ is about 0 for $\mathcal{K} < 17$ and

3.8 for $\mathcal{K} > 17$ with small differences depending on the surface orientations, which exactly correspond to the anisotropy of surfaces.

The physical origin of the slope $k$ likely stems from the ratio term $CN/\overline{CN}$ (where $CN$ and $\overline{CN}$ are the usual coordination number and the generalized coordination number[29,30] of surface atoms) with Supplementary Eq. 1. $CN/\overline{CN}$, which is an indicator of the atomic packing density of surface atoms, provides an effective measure to combine together the surfaces across the different crystals such as fcc, bcc and hcp. For the CPSs of fcc, bcc and hcp (that are (111), (110) and (0001) surfaces), $CN/\overline{CN}$ is 6/5, generating the slope $k = -0.033$ for crystals with $\mathcal{K} > 17$ and $k = 0.20$ for crystals with $\mathcal{K} < 17$. The coordination number and slope values for the other surface groups as well as the reason for grouping are summarized in Supplementary Note 1 and Supplementary Tables 3-5. The predicted slopes $k$ by Supplementary Eq. 1 are in agreement with the fits of the DFT-calculated and experimental results for elemental solids[14,26–28], and can be approximated as constant for $\mathcal{K} > 17$ or $\mathcal{K} < 17$.

Although the scaling rule is apparent, there exists some outliers deviating from the guiding lines for the experimental results, e.g. the CPSs of Cr and Mn in Fig. 1f and i and for the calculated results, e.g. on some high-index surfaces of Tl, Os and Be in Supplementary Figs. 1 and 2. The deviations are attributed to the potential phase transition in experiments (fcc-bcc for Cr and Mn[26]) and the possible insufficient description of surface reconstruction in calculations. It is known that Be exhibits unexpected expansion compared to the other metals[31] and has been demonstrated as a reflection of the novel surface electronic states such as the anomalous surface electron-phonon coupling[32,33].

Since the experimental surface energy for solids is obtained by extrapolating from liquid-phase measurements[34–41], we also study surface energies for liquid phases of the considered systems, by considering the coordination difference between solid and liquid phases (see the details in Supplementary Note 1), as $\gamma = \left(1 - \sqrt{\frac{|CN_s - CN_l|}{CN_s}}\right) k\mathcal{K} + b$, where $CN_l$ and $CN_s$ are the coordination number of bulk liquid and solid phases respectively. The correlation between surface energies of liquids and the electronic descriptor $\mathcal{K}$ is shown in Fig. 1i and Supplementary Tables 1, 2 and 5. The slopes of the fitted linear relation are 0.150 for $\mathcal{K} < 17$ and -0.022 for $\mathcal{K} > 17$, which are significantly different from those of solids and are in good agreement with the predictions by our scheme.

**The anisotropy of surface energy for solids.** Here we turn to understand the anisotropy of surface energy, by studying three kinds of crystal structures with 37 different surfaces[28]. We find that the surface energies on each solid exhibit a linear relationship with the generalized coordination number $\overline{CN}$ (see Fig. 2a-c and Supplementary Figs. 4 and 5), as,

$$\gamma = \lambda\overline{CN} + \xi \quad (3)$$

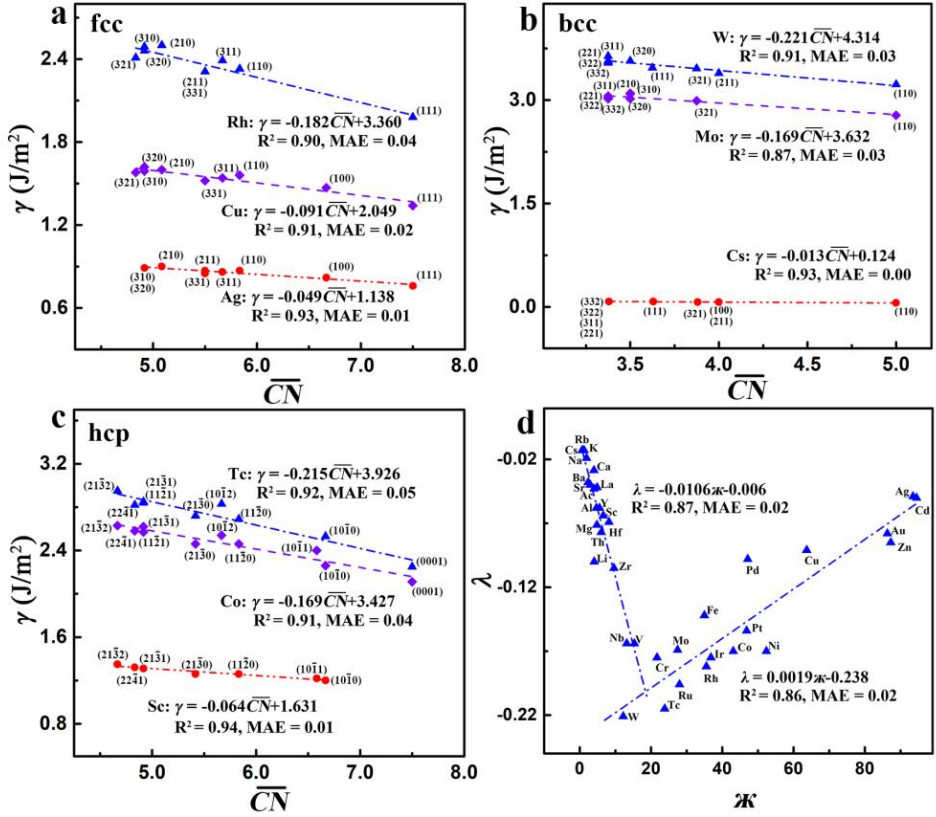

**Figure 2 | The anisotropy of surface energy for solid metals.** a, b, c, DFT-calculated surface energies against the generalized coordination number $\overline{CN}$ for fcc metals on Cu, Ag and Rh surfaces[28] (a), for bcc metals on Mo, Cs and W surfaces[28] (b), and for hcp metals on Sc, Tc and Co surfaces[28] (c). d, The slope $\lambda$ of the $\overline{CN}$ determined scaling relation against the electronic descriptor $\mathcal{K}$ for fcc, bcc and hcp metals[28].

where the $\lambda$ and $\xi$ are constant for a given solid. We further demonstrate that $\lambda$ scales linearly with the electronic descriptor $\mathcal{K}$ (Fig. 2d), approximately as,

$$\lambda = \begin{cases} -\dfrac{1}{100}\mathcal{K}, & \mathcal{K} < 17 \\ \dfrac{1}{500}\mathcal{K} - \dfrac{1}{4}, & \mathcal{K} > 17 \end{cases} \quad (4)$$

Clearly, the predicted prefactors by Eq. (4) are in good agreement with the fits of DFT-calculated results[28] in Fig. 2d. Our scheme allows one to understand deeply the anisotropy of surface energy for different solids. For solids with large or small $\mathcal{K}$ such as Cu, Au, Ag, Na, K and Ca, the prefactor $\lambda$ is small, leading to the weak anisotropy of surface energy for these solids. In contrast, for solids with medium $\mathcal{K}$ such as Rh, Mo, Co and Tc, the large prefactor term $\lambda$ makes the anisotropy of these solids become greater. All these predictions by Eq. (4) are in accordance with the DFT findings in Fig. 2 and Supplementary Figs. 4 and 5[28].

By correlating $\xi$ with the electronic descriptor $\mathcal{K}$, we obtain a linear scaling between $\xi$ and $\mathcal{K}$ as Eq. (5),

$$\xi = \begin{cases} \dfrac{6}{25}\mathcal{K}, & \mathcal{K} < 17 \\ -\dfrac{6}{125}\mathcal{K} + \mu, & \mathcal{K} > 17 \end{cases} \quad (5)$$

**A scheme combining material-dependent property and anisotropy of surface energy for solids**. We thus propose an entire expression of surface energy for solids based on the electronic descriptor $\mathcal{K}$ and the geometric descriptor $\overline{CN}$ as:

$$\gamma = \begin{cases} \dfrac{1}{100}\times(24-\overline{CN})\mathcal{K}, & \mathcal{K} < 17 \\ -\dfrac{1}{500}\times(24-\overline{CN})\mathcal{K} - \dfrac{1}{4}\overline{CN} + \mu, & \mathcal{K} > 17 \end{cases} \quad (6)$$

where $\mu$ is a constant. This equation quantitatively describes the material-dependent nature and anisotropy of surface energy for solids, and can be used to estimate rapidly the trends of surface energy on different materials and surfaces, as the involved parameters are easily accessible (see more details in Supplementary Note 1).

**Generalize the model into alloys and semiconductor compounds.** Alloying can generate much complex bulk structures and/or coordination environments than elemental crystals. Body-centered-tetragonal intermetallics is an important family of binary alloys with formula AB, where elements A and B locate at the vertex and the center of the cube. Fluorite-structure intermetallics is one of the important binary alloys with formula $A_2B$. This kind of alloys has a typical fluorite-like structure where the TM element B is arranged in fcc stacking and the tetrahedral interstice is filled with the main-group element A. Face-centered-cubic intermetallics also plays vital role in alloys with formula $A_3B$, in which elements A and B locate at the face center and vertex of the cubic respectively. Mg-based surface alloys are constructed by substituting one Mg atom at the surface with another element M, which can significantly modulate the properties of Mg surfaces. III-V semiconductor compounds are typical Zinc blende structure with cation (such as Al, Ga and In) forming fcc structure and anion (such as P, As, Sb and Bi) filling in the tetrahedral interstice. Each of these four crystal structures exhibits variable coordination environments from one surface orientation to the next.

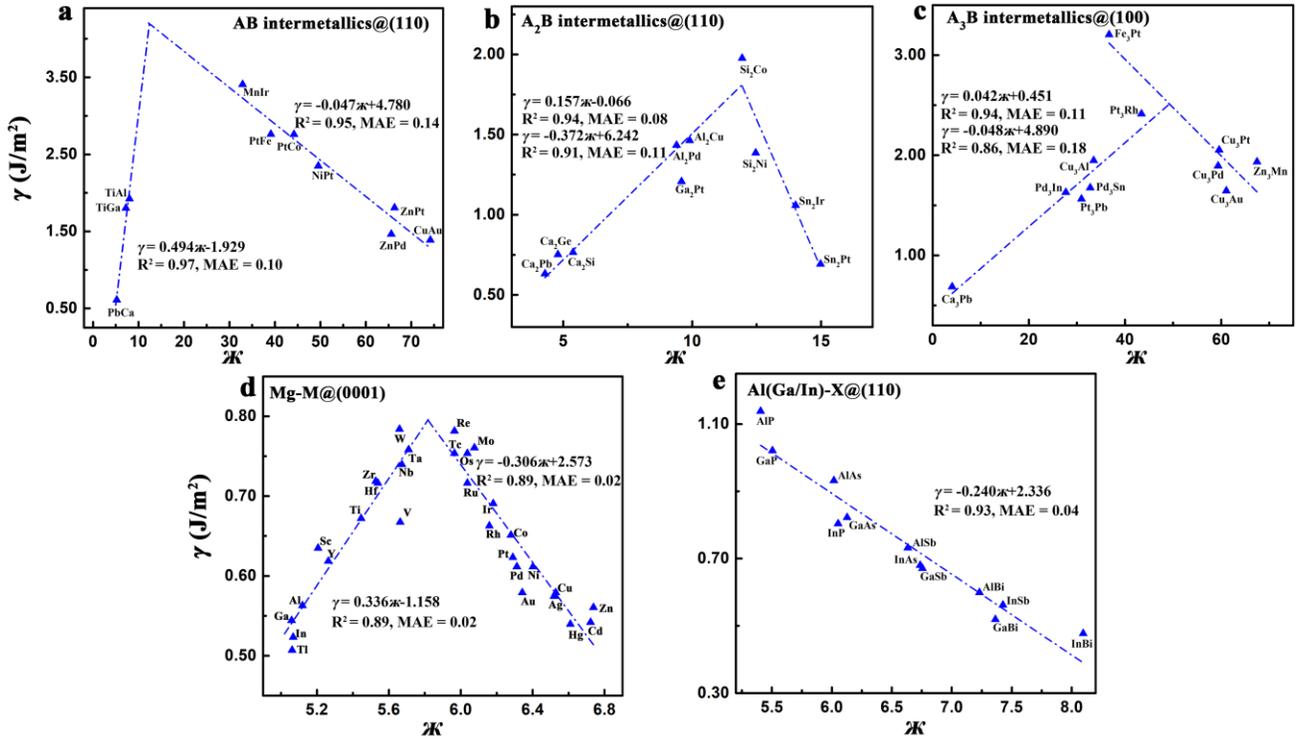

**Figure 3 | Cleavage energies and surface energies of alloys and semiconductor compounds as a function of the electronic descriptor $\mathcal{K}$.** a, Cleavage energies for (110) surface of body-centered-tetragonal (AB) intermetallics[42]. b, Cleavage energies for (110) surface of fluorite-structure (A₂B) intermetallics[42]. c, Cleavage energies for (100) surface of face-centered-cubic (A₃B) intermetallics[42]. d, Surface energies for (0001) surface of Mg-based (Mg-M) surface alloys[43]. e, Surface energies for (110) surface of Al series (AlP, AlAs, AlSb, AlBi), Ga series (GaP, GaAs, GaSb, GaBi) and In series (InP, InAs, InSb, InBi) semiconductor compounds[44].

It is encouraging that our established correlation can be readily generalized into AB intermetallics, A₂B intermetallics, A₃B intermetallics, Mg-based surface alloys, and semiconductor compounds by considering the environment effect of bulk and surface atoms. For bulk atoms, the period number $N_p$ and group number $N_g$ are obtained by averaging the period and group number based on the stoichiometric ratio between A ($n_A$) and B ($n_B$), that is: $\left(N_{p,A}^{n_A} N_{p,B}^{n_B}\right)^{1/(n_A+n_B)}$ and $\left(N_{g,A}^{n_A} N_{g,B}^{n_B}\right)^{1/(n_A+n_B)}$. For surface atoms, $\psi$ is obtained as $\frac{\left(\prod_{i=1}^{N} S_{vi}\right)^{2/N}}{\left(\prod_{i=1}^{N} \chi_i\right)^{1/N}}$, where $S_{vi}$ and $\chi_i$ are the numbers of outer electrons and the electronegativity of the $i$th atom around the active surface center, and $N$ is the sum number of atoms around the active surface center (see more details in Supplementary Note 2). Note that this $\mathcal{K}$ expression will naturally transform into Eq. (1) in calculating elemental crystals and is thus universal for elemental crystals, alloys and compounds.

We study cleavage energies for 11 types of AB intermetallics with 12 different orientations, for 11 types of A₂B intermetallics with 12 different orientations and for 13 types of A₃B intermetallics with 3 different orientations, and surface energies for 31 Mg-M surface alloys with (0001), (10$\bar{1}$0) and (11$\bar{2}$0) surfaces, and for 12 types of semiconductor compounds with (110) surfaces[42-

[44]. The corresponding $\mathcal{K}$ values are obtained in Supplementary Note 2 and are summarized in Supplementary Tables 6-10. The $\overline{CN}$ values of all surfaces for intermetallics are summarized in Supplementary Table 11.

The cleavage energies of AB intermetallics, A₂B intermetallics and A₃B intermetallics exhibit a linear relation with the electronic descriptor $\mathcal{K}$ (see Fig. 3a-c and Supplementary Figs. 6-9)[42]. Encouragingly, the scaling relations can be expressed with $\gamma=\frac{1}{100}(32-\overline{CN})\mathcal{K}$ for $\mathcal{K} < 15$ and $\gamma=-\frac{1}{500}(24-\overline{CN})\mathcal{K}-\frac{1}{4}\overline{CN}+5.7$ for $\mathcal{K} > 15$ of AB intermetallics, $\gamma=\frac{1}{100}(24-\overline{CN})\mathcal{K}$ for $\mathcal{K} < 12$ and $\gamma=-\frac{1}{500}(225-\overline{CN})\mathcal{K}-\frac{1}{4}\overline{CN}+8.2$ for $\mathcal{K} > 12$ of A₂B intermetallics, and $\gamma=\frac{1}{100}(12-\overline{CN})\mathcal{K}$ for $\mathcal{K} < 50$ and $\gamma=-\frac{1}{500}(24-\overline{CN})\mathcal{K}-\frac{1}{4}\overline{CN}+5.62$ for $\mathcal{K} > 50$ of A₃B intermetallics, implying that our scheme can also characterize the materials-dependent properties and anisotropy of cleavage energy for these alloys.

For Mg-M surface alloys, the proportion of alloying element ($\eta$) is a non-negligible factor that controls how the alloys are formed, thereby affects surface energy, as

$$\gamma = \begin{cases} (6.0\eta - 0.27) \times \mathcal{K} + b_1, & \mathcal{K} < 5.9 \\ (6.1\eta - 0.91) \times \mathcal{K} + b_2, & \mathcal{K} > 5.9 \end{cases} \quad (7)$$

for (0001), (10$\bar{1}$0) and (11$\bar{2}$0) surfaces, $\eta$ is 1/10, 1/9 and 1/8 respectively. Overall, Eq. (7) successfully draws the trends of surface energies for Mg-M alloys on different surfaces (Fig. 3d, Supplementary Fig. 10 and Supplementary Table 12)[43]. Fig. 3e shows that the surface energies of semiconductor compounds also exhibit a linear relationship with $\mathcal{K}$[44] with the similar scaling rule as that for elemental crystals $\gamma = -k\mathcal{K} + b$, where the prefactor $k$ for semiconductor compounds is -0.24. These results demonstrate that our scheme indeed captures the intrinsic properties of cleavage energy and surface energy and is thus universal in describing the surface stability of elemental crystals, alloys and semiconductor compounds.

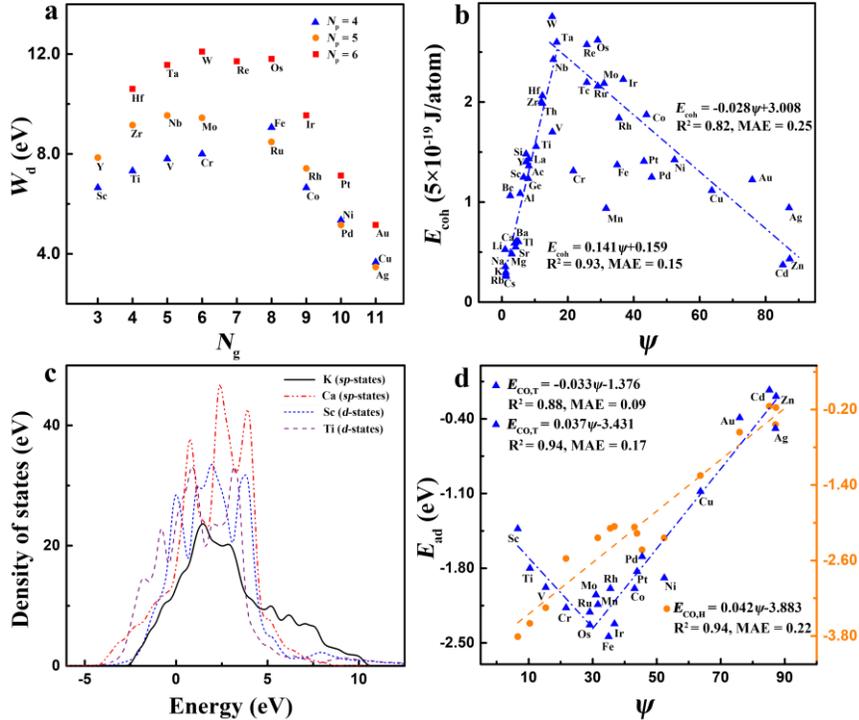

**Figure 4 | Electronic structures and energetics of elemental crystals.** a, The correlation between the $d$-band width ($W_d$) and the period number ($N_p$) and group number ($N_g$) for 3$d$-, 4$d$- and 5$d$-series transition metals (TMs)[25]. b, Experimental cohesive energies of elemental crystals against the electronic descriptor $\psi$[46]. c, Density of states of the $sp$-bands of K and Ca atoms and the $d$-bands of Sc and Ti atoms on close-packed surfaces. d, CO adsorption energies against the electronic descriptor $\psi$ on close-packed surfaces of TMs at top (T) and hollow (H) sites.

**Understanding and progress of the model.** To deeply understand the electronic descriptor $\mathcal{K}$, we analyze the role of the involved parameters ($N_p$, $N_g$, $S_v$ and $\chi$) by taking TMs as an example. Fig. 4a shows the $d$-band width that is usually taken as an index of bond energy, with respect to the period number $N_p$ and group number $N_g$[25]. In each period, the $d$-band width scales with $N_g$ in a broken-line behavior from group 3 to group 11: it first increases until groups 6~8 and then decreases. In each group, the $d$-band width increases with increasing the period number for groups 3~6, whereas this proportional relationship is broken for groups 8~11. These results indicate the $d$-band width is linearly related to $N_g$ in each period and is correlated with $N_p$ depending on $N_g$. According to the tight-binding (TB) approximation, the energy gain of forming surface is thus scaled with $(N_p)^{1/2}$ because of the downshift of the occupied states[45], while the $N_g$-dependent term likely enters the exponential term of $N_p$. This corresponds to the relation of $\gamma \propto N_p \sqrt{N_g}$, as we found. In addition, $S_v$ and $\chi$ (forming $\psi$) provides a reasonable description of cohesive energy (see Fig. 4b)[46]. As a result, the electronic descriptor $\mathcal{K}$, which combines $N_p$, $N_g$, $S_v$ and $\chi$, can describe surface energy and cleavage energy effectively.

Why the early TMs (groups 3-5) are consistent with the main-group crystals instead of the late TMs in surface energy. Compared with the late TMs, the $d$-bands of the early TMs exhibit the centers above Fermi energy level and the large width, which are close to the $sp$-bands of the main-group crystals (see Fig. 4c) rather than the $d$-bands of the late TMs. Looking from practical effect, the early TMs thus behave similarly with the main-group crystals in surface energy.

We treat that $\overline{CN}$ is a better descriptor in describing the surface-energy anisotropy of solids compared with $CN$. While $\overline{CN}$ and $CN$ are similar in characterizing the surface-energy anisotropy of fcc and hcp crystals, $\overline{CN}$ performs better than $CN$ for bcc crystals, with $R^2$ much larger for $\overline{CN}$ (>0.90) than for $CN$ (~0.80) (see Fig. 2 and Supplementary Figs. 4, 5 and 11). $CN$ is the number of the nearest neighbors for a given surface atom, while $\overline{CN}$ considers both the first- and second-nearest neighbors[29,30,47]. Therefore, $\overline{CN}$ captures more general geometry of surfaces and is more sensitive to the variation of different orientations, thereby providing a more accurate description for the anisotropy of surfaces.

We now do a brief comparison between our model and the conventional models such as the broken-bond models[11-13], Miedma model[14], Friedel model[15] and Stefan model[15]. First, the conventional models connect surface energy with another energies or $d$-band width, which require expensive experiments or calculations when being applied into alloys, while our model correlates surface energy with the intrinsic properties of surfaces, including the period number and group number of bulk atoms, and the valence-electron number, electronegativity and coordination of surface atoms, all of which are easily accessible by table lookup. Secondly, the conventional models are generally only applicable into elemental crystals, while our model is effective for the elemental crystals in both solid and liquid phases, AB intermetallics, $A_2B$ intermetallics, $A_3B$ intermetallics, Mg-based surface alloys and semiconductor compounds. Third, compared with the experimental results for elemental crystals, the mean absolute error (MAE) is 0.47 J/m² for the bond-cutting model, 0.50 J/m² for the square-root bond cutting model, 1.40 J/m² for the Friedel model and 4.58 J/m² for the Stefan model, whereas the MAE of our model by Eq. (6) is 0.28 J/m² (see Supplementary Table 13). For the alkaline metals and alkaline-earth metals that have relatively small surface energies, the MAE of our scheme is only 0.12 J/m². When estimating surface energy on 45 elemental crystals, 66 alloys, and 12 semiconductor compounds (totally 884 different values), the MAE of the predictions by our model relative to the DFT calculations is ~0.23 J/m² (see the sheet of "Predicted surface energies" in Source Data file), which is about 0.16 eV/atom, smaller than the approximate error of semi-local functionals. Last but not least, the conventional models can hardly reveal the anisotropy of surface energy, while our model can well characterize this effect. Our model thus provides a more explicit and accurate physical picture for the surface stability of materials (see more details in Supplementary Note 3).

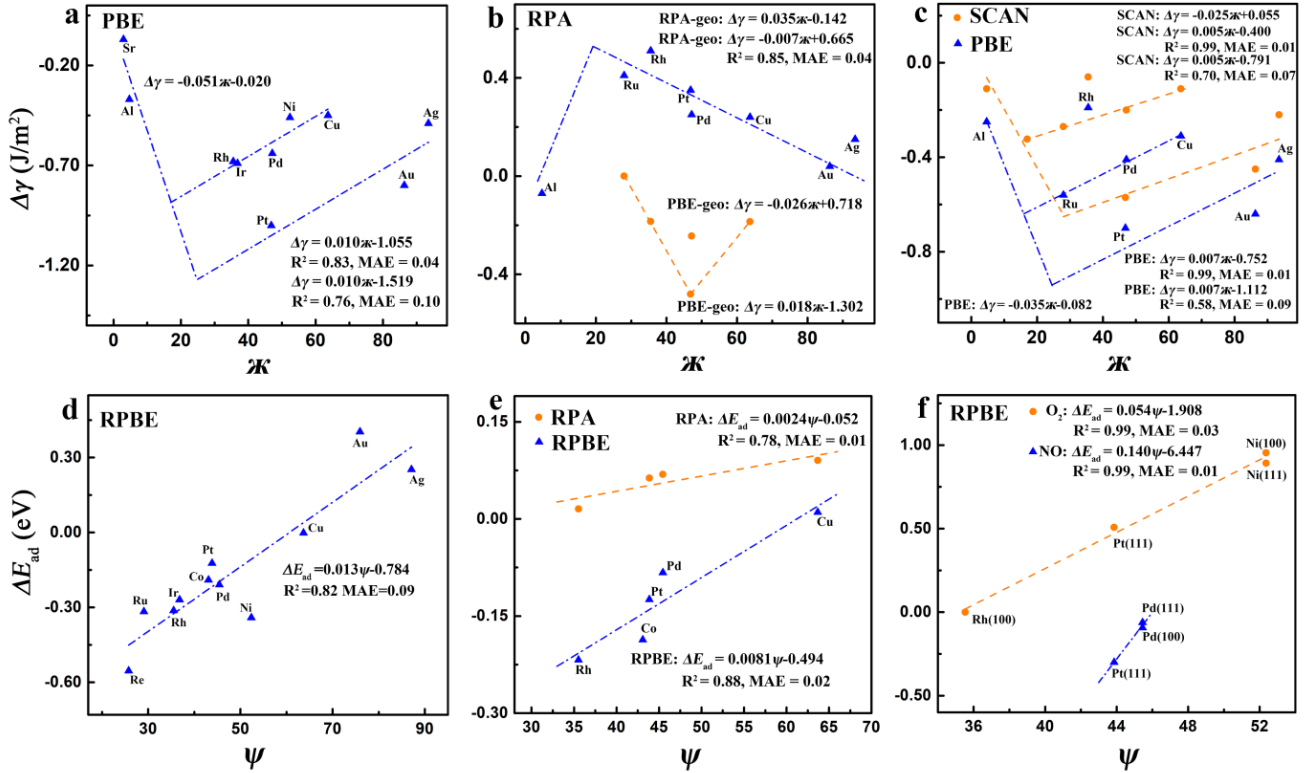

**Figure 5 | The errors of first-principle methods in calculating surface energy and adsorption energy against the electronic descriptors 𝒦 and ψ.** a, The error of surface energies calculated by PBE functional in fcc(111) surfaces[28]. b, The error of surface energies calculated by RPA functional with RPA geometries[19] (blue triangle) and PBE geometries[10] (orange circle) in fcc(111) surfaces. c, The error of surface energies calculated by PBE functional[19] (blue triangle) and SCAN functional[19] (orange circle) in fcc(111) surfaces. d, The error of CO adsorption energy calculated by PBE functional on transition-metal (TM) close-packed surfaces[17]. e, The error of CO adsorption energy calculated by PRBE functional[18] (blue triangle) and RPA functional[10] (orange circle) on TM close-packed surfaces. f, The error of adsorption energy of NO (blue triangle), and $O_2$ (orange circle) calculated by PRBE functional[18] on TM surfaces.

Wulff shape plays a vital role in understanding surface properties of crystals especially those of nanoparticles that have relatively large surface areas[48]. Our scheme also allows the fast estimation of surface energy for Wulff shape, with the relation of $\bar{\gamma} = \sum_{\{hkl\}} \gamma_{hkl} f_{hkl}^A$, where $f_{hkl}^A$ denotes the area fraction of a given facet $\{hkl\}$ on the Wulff shape. We consider the Wulff shape for 40 elemental crystals and the maximum Miller index of all surfaces is up to 3 (the higher-index surfaces are indispensable for the equilibrium of crystals). The small MAE (~0.19 J/m²) of our predictions compared with the DFT-calculated results[28] supports the effectiveness of our model in the future design of Wulff shape (see Supplementary Note 4 and Supplementary Table 14).

**The quantitative correlation between surface energy and adsorption energy.** We now try to uncover the quantitative correlation between surface energy and adsorption energy. In addition to the expression of surface energy with Eq. (6), we had also demonstrated a general expression for adsorption energy $E_{ad}$ on metallic materials as follows[23],

$$E_{ad} = 0.1 \times \frac{X_m \cdot X}{X_m + 1} \times \psi + 0.2 \times \frac{X+1}{X_m+1} \times \overline{CN} + \theta \quad (8)$$

where $X$ and $X_m$ are the actual bonding number and maximum bonding number of the central atom for a given adsorbate. $\theta$ is a constant originating from the coupling between the adsorbate states and the substrate-$sp$ states. Adsorption energies of CO, NO and O are shown in Fig. 4d and Supplementary Fig. 12, where the fitted prefactors are consistent with the predictions by Eq. (8)[49,50] (see Supplementary Note 5). Combining Eqs. (6) and (8), we build a quantitative relation between surface energy and adsorption energy,

$$E_{\text{ad}} = \tag{9}$$

$$= \begin{cases} \dfrac{10\gamma}{(24-\overline{CN})} \times \left(\dfrac{N_p}{4}\right)^{(3-\sqrt{N_g})} \times \dfrac{X_m-X}{X_m+1} + 0.2 \times \dfrac{X+1}{X_m+1} \times \overline{CN} + \theta & \mathscr{K} < 17 \\ -\dfrac{50\gamma+12.5\overline{CN}\cdot50\mu}{(24-\overline{CN})} \times \left(\dfrac{N_p}{4}\right)^{(3-\sqrt{N_g})} \times \dfrac{X_m-X}{X_m+1} + 0.2 \times \dfrac{X+1}{X_m+1} \times \overline{CN} + \theta, & \mathscr{K} > 17 \end{cases}$$

Surface energy is thus correlated with adsorption energy in both positive and negative relationship, dismissing the conventional concept that $E_{\text{ad}}$ is positively correlated with $\gamma$. Taking CO adsorption as an example, the underlying mechanism can be understood by the different responses of $5\sigma$- and $2\pi*$-metal hybrid orbitals at the different adsorption sites[51-53] (see the details in Supplementary Note 6). It is noteworthy that surface energy is approximately linearly scaled with adsorption energy for the late TMs, as their term is approximately constant.

Our scheme provides a simple picture for understanding the connections and distinctions between surface energy and adsorption energy [see Eq. (9)]. While the valence-electron number, electronegativity and coordination of surface atoms are crucial for both surface energy and adsorption energy, the period number and group number of bulk atoms play an additional decisive role for surface energy. For surface energy, a remarkable characteristic is that the effects of electronic and geometric properties are coupled, leading to a coupling term $\mathscr{K}\overline{CN}$ ($\left(\dfrac{N_p}{4}\right)^{(\sqrt{N_g}\cdot3)} \times \psi\overline{CN}$) [see Eq. (6)]. For adsorption energy, the electronic and geometric properties ($\psi$ and $\overline{CN}$) are generally independent of each other, while these two determinants are coupled with the valence of adsorbates respectively [see Eq. (8)]. Moreover, our scheme also allows the estimation of adsorption energy using surface energy with Eq. (9). To justify the prediction accuracy, we adopt the fcc(111) surfaces of TMs and the adsorbate CO with both experimental and theoretical results[19,20] (see Supplementary Note 7 and Supplementary Table 15). The MAEs of the predicted adsorption energies by Eq. (9) relative to the experimental and calculated results is 0.08 eV for experiments, 0.16 eV for local density approximation (LDA)[54], 0.17 eV for Perdew−Burke−Ernzerhof (PBE)[55], 0.15 eV for PBEsol (PBE for solids)[56], and 0.23 eV for strongly constrained and appropriately normed (SCAN)[57] meta-generalized gradient approximation (GGA), most of which are smaller than the $\pm 0.2$ eV, the approximate error of DFT (semi-)local functionals. These consistencies demonstrate that our established relation between surface energy and adsorption energy is robust and can be applied into the future materials design.

## Discussion

We now study the origin of the material-dependent error of first-principle methods in calculating surface energy and adsorption energy.

For surface energy, the difference between the DFT-calculated and experimental results can be expressed as,

$$\Delta\gamma = \gamma' - \gamma \tag{10}$$

$$= \begin{cases} -\dfrac{\Delta_2}{100}\mathscr{K} + \dfrac{1}{100} \times (24-\overline{CN}\cdot\Delta_2)\Delta_1, & \mathscr{K} < 17 \\ \dfrac{\Delta_2}{500}\mathscr{K} - \dfrac{1}{500}(24-\overline{CN}\cdot\Delta_2)\Delta_1 - \dfrac{1}{4}\Delta_2 + \Delta\mu, & \mathscr{K} > 17 \end{cases}$$

For simplicity, we first assume that $\Delta_1 = \mathscr{K}' - \mathscr{K}$ and $\Delta_2 = \overline{CN}' - \overline{CN}$ are both constant for a given method, corresponding to the systematical error of first-principle calculations ($\mathscr{K}'$ and $\overline{CN}'$ denote the calculated electronic and geometric structures). $\Delta\gamma$ is thus dominated by the descriptors $\mathscr{K}$ for a given method. Eq. (10) deduces that $\Delta\gamma$ should exhibit a material-dependent nature with an approximately linear scaling with $\mathscr{K}$ and the corresponding slope, mainly determined by the geometric error $\Delta_2$, for $\mathscr{K} < 17$ is about five times of that for $\mathscr{K} > 17$. These deductions have been explicitly demonstrated by the calculations with LDA, PBE, PBEsol, SCAN, SCAN+rVV10 (Vydrov−Van Voorhis 2010[58]), and RPA (see Fig. 5a-c and Supplementary Fig. 13)[10,19,28]. Notably, RPA calculations with RPA geometries overestimate but RPA with PBE geometries underestimate surface energy, again identifying the key role of geometric error $\Delta_2$ (see Fig. 5b and Supplementary Note 8)[10,19]. As these six methods can describe the electronic structures of metals with reasonable accuracy, the intercept for $\mathscr{K} < 17$ is small, indicating that the metals with small or large $\mathscr{K}$ ($\mathscr{K} < 10$ or $\mathscr{K} > 60$) likely generate smaller $\Delta\gamma$ than those with medium $\psi$ ($10 < \mathscr{K} < 60$). This again has been demonstrated by the first-principle methods that $\Delta\gamma$ is much smaller in Al, Sc, Au and Ag than in Pt, Rh and Re (see Fig. 5a-c and Supplementary Fig. 13)[27,28]. These results strongly support the robustness of our model for surface energy.

In the case of adsorption energy, the error of adsorption energy between first-principle calculations and experiments is as follows,

$$\Delta E_{\text{ad}} = \dfrac{-\Delta_3}{10(X_m+1)} \times \psi + \dfrac{2\overline{CN}\Delta_3+(X_m-X')\Delta_4+2(X'+1)\Delta_2}{10(X_m+1)} + \theta_{1,2} \tag{11}$$

$\Delta_3 = X' - X$ and $\Delta_4 = \psi' - \psi$ are the calculated error of electronic structures of adsorbates and substrates ($X'$ and $\psi'$ denote the calculated electronic structures of adsorbates and substrates). Eq. (11) predicts that $\Delta E_{\text{ad}}$ should be a linear function of the electronic descriptor $\psi$ with the slope of $\dfrac{-\Delta_3}{10(X_m+1)}$. Indeed, the calculations by LDA, Perdew-Wang-91 functional (PW91)[59], PBE, PBEsol, Bayesian error estimation functional with van der Waals correlation (BEEF-vdW)[60], Hammer, Hansen, Nørskov modified PBE functional (RPBE)[61], and RPA fulfill this prediction, generating linear relations between $\Delta E_{\text{ad}}$ and $\psi$ for CO, NO, and $O_2$ adsorption on TMs (see Fig. 5d-f and Supplementary Fig. 14)[10,17,18]. Compared to LDA, GGAs and meta-GGAs, RPA exhibits a smaller $\Delta_3$ (-0.5 versus -0.1 for CO), suggesting that the material-dependent error is small for RPA in describing adsorption energy. For CO adsorption, the constant $\theta_{1,2}$ term in Eq. (11) is most likely larger for GGAs and meta-GGAs than for RPA, as GGAs and meta-GGAs underestimate significantly the energy of $2\pi^*$ orbits of CO[9,10]. Due to the synergy of $0.1\Delta_3\psi/(\mathcal{X}_m+1)$ and $\theta_{1,2}$, RPBE, PBE and SCAN overestimate the adsorption energy for metals with small $\psi$ such as Pt and Rh but underestimate the adsorption energy for metals with large $\psi$ such as Ag (see Fig. 5d and Supplementary Fig. 14b)[17,20], whereas RPA underestimate slightly the adsorption energy on all considered TMs (see Fig. 5e)[10] (see more details in Supplementary Note 7).

Overall, our scheme indicates that the observed material-dependent behavior of the first-principle methods in calculating surface energy and adsorption energy is due to the coupling between the constant systematic errors and the electronic properties of materials denoted by $\mathcal{K}$ and $\psi$, in which the electronic structure $\mathcal{K}$ and $\psi$ strongly depend on materials, as shown in Eqs. (10) and (11).

We have proposed an effective model for the accurate determination of surface energy based on the period number and group number of bulk atoms, and the valence, electronegativity and coordination of surface atoms, which holds for main- and transition-group elemental crystals in both solid and liquid phases, AB intermetallics, $A_2B$ intermetallics, $A_3B$ intermetallics, Mg-based surface alloys and semiconductor compounds. We find that the electronic properties of bulk and surface atoms control the material-dependent property of surface energy, while the electronic properties of bulk and surface atoms and generalized coordination of surface atoms jointly determine the anisotropy of surface energy. This model correlates surface energy with the intrinsic properties of materials, builds the quantitative correlation between surface energy and adsorption energy, and uncovers the origin for the material-dependent error of first-principle methods in calculating surface energy and adsorption energy. Our findings rationalize some theoretical and experimental results, and could be helpful to engineering surface energy and adsorption energy simultaneously for materials design.

## Methods

In the study, the density of states (DOS) calculations were performed by CASTEP code[62] with ultrasoft pseudopotentials[63] and PBE functional augmented with TS method[64]. TM surfaces were modeled with four-layer slabs in a unit cell of p(2×2), where the top two layers were fully relaxed and the rest of the layers were constrained in the optimized lattice. A vacuum of 15 Å was adopted to separate the adjacent slabs. We used plane-wave cutoff energy of 450 eV and the Monkhorst-Pack $k$-point sampling with 8×8×2 meshes for geometry optimization. The conjugate gradient algorithm was utilized with a convergence threshold of $5.0e^{-6}$ eV and 0.01 eV/Å in Hellmann-Feynman force on each atom. It is noteworthy that only the DOS of TMs were calculated by our PBE+TS methods, while the rest data such as surface energies, cleavage energies and adsorption energies were cited from literatures[14,26–28,34–44,49,50]. Surface energy and cleavage energy are calculated with the slab models[26,28,42-44] as,

$$\gamma = \frac{E_{slab} - nE_{bulk}}{2A} \quad (12)$$

where $E_{slab}$ is the energy of the slab, $E_{bulk}$ is the bulk energy per atom, $n$ is the number of the atoms in the slab, and A is the surface area of the slab. Eq. (12) corresponds to surface energy, when cleavage of bulk yields symmetric slabs[26,28,43,44], e.g. for elemental crystals, which have two identical surface terminations and equal to surface energy. If cleavage of bulk generates asymmetric slabs, e.g. for intermetallic compounds, which have two distinct terminations and different surface energy, Eq. (12) obtains cleavage energy[42], which is the average of the surface energy of the two different termination. Note that cleavage energy is identical to surface energy only when the slabs are symmetric. Encouragingly, our scheme works well for both surface energy and cleavage energy.

For Mg-based surface alloys, the surface slabs were constructed by substituting one Mg surface atom with another alloying element M on each termination of the slab to ensure the symmetry. The $Mg_{n-2}M_2$ bulk system contains 94 Mg atoms and 2 M atoms ($n$ is large enough) so that the distance between the 2 M atoms is large enough, hence the interaction between the 2 M atoms can be ignored. Then surface energies of Mg-M surface alloys[43] are calculated as,

$$\gamma = \frac{E_{slab}[(m\text{-}2)Mg + 2M] - E_{bulk}(Mg_{n\text{-}2}M_2) - (m\text{-}n)E_{Mg}}{2A} \quad (13)$$

where $E_{slab}[(m\text{-}2)Mg + 2M]$ and $E_{bulk}(Mg_{n\text{-}2}M_2)$ are the total energy of the Mg-M slabs and Mg-M bulk phase. $E_{Mg}$ represents the energy of Mg per atom in the bulk hcp phase and A is the surface area of the slab.

## Author information

*E-mail: wgao@jlu.edu.cn
*E-mail: jiangq@jlu.edu.cn

## Author contributions

W. G. and Q. J. conceived the original idea and designed the strategy. W. G. derived the models and analyzed the results with the contribution from B. L. and X. L. W. G. and B. L. wrote the manuscript. B. L. prepared the Supplementary Information and drew all figures. All authors have discussed and approved the results and conclusions of this article.

## Acknowledgements

The authors are thankful for the support from the Program of the Thousand Young Talents Plan, the National Natural Science Foundation of China (Nos. 21673095, 11974128, 51631004), the Opening Project of State Key Laboratory of High Performance Ceramics and Superfine Microstructure (SKL201910SIC), the Program of Innovative Research Team (in Science and Technology) in University of Jilin Province, the Program for JLU (Jilin University) Science and Technology Innovative Research Team (No. 2017TD-09), the Fundamental Research Funds for the Central Universities, and the computing resources of the High Performance Computing Center of Jilin University, China.

## Competing interests

The authors declare no financial competing interests.